\documentclass[journal]{IEEEtran}
\usepackage{amsmath}
\newtheorem{definition}{Definition}

\newtheorem{proposition}{Proposition}
\newtheorem{lemma}{Lemma}

\newtheorem{corollary}{Corollary}

\begin{document}
%\title{Properties of Quantum Spectral Information Rates}
\title{Beyond {i.i.d.} in Quantum Information Theory}
\author{Garry~Bowen and Nilanjana~Datta% <-this % stops a space
\thanks{This work was supported by the EPSRC (Research Grant GR/S92816/01).}%
\thanks{G. Bowen is with the Centre for Quantum Computation, Department of Applied Mathematics and Theoretical Physics, University of Cambridge, Cambridge CB3 0WA, UK (e-mail: gab30@damtp.cam.ac.uk).}
\thanks{N. Datta is with the Statistical Laboratory, Department of Pure Mathematics and Mathematical Statistics, University of Cambridge, Cambridge CB3 0WA, UK (e-mail: N.Datta@statslab.cam.ac.uk).}%
%} <-this % stops a space
}
\markboth{Submitted to IEEE Transaction on Information Theory}{Bowen \MakeLowercase{\textit{et al.}}: Beyond {i.i.d.} in Quantum Information Theory}
% *** Note that you probably will NOT want to include the author's name in ***
% *** the headers of peer review papers.                                   ***

% If you want to put a publisher's ID mark on the page
% (can leave text blank if you just want to see how the
% text height on the first page will be reduced by IEEE)
% \pubid{0000--0000/00\$00.00~\copyright~2002 IEEE}

%\specialpapernotice{(Invited Paper)}

\maketitle

\begin{abstract}
The information spectrum approach gives general formulae for optimal rates of codes in many areas of information theory.  In this paper the quantum spectral divergence rates are defined and properties of the rates are derived.  The entropic rates, conditional entropic rates, and spectral mutual information rates are then defined in terms of the spectral divergence rates.  Properties including subadditivity, chain rules, Araki-Lieb inequalities, and monotonicity are then explored.
\end{abstract}


% Will be ignored for other modes.
\IEEEpeerreviewmaketitle

\section{Introduction}

\PARstart{T}{raditionally,} rates for data compression, channel capacity, and other
operational quantities in information theory are related to entropic functions
of the state (or distribution).  The underlying assumption is that the source
or channel is identical and independently distributed ({i.i.d.}), or
memoryless, over many uses, and the rate is determined in the asymptotic limit.
The entropies themselves obey various mathematical relationships, many of which
have additional operational interpretations.

The information spectrum approach of Han \& Verdu gives asymptotic rate
formulae for many operational schemes in information theory, such as data
compression, channel capacity, and hypothesis testing, where very few
assumptions are made about the source or channel \cite{verdu94,han}.

In quantum information theory the ideas of coding and communication are
generalized to include the nature of the physical system in which information
is encoded.  Spectral information rates for quantum states were derived by
Ogawa, Hayashi \& Nagaoka, initially in terms of hypothesis testing and source
coding \cite{ogawa00,nagaoka02}, and additionally to determine general
expressions for entanglement concentration \cite{hayashi03a}, and the classical
capacity of arbitrary quantum channels \cite{hayashi03}.

In this paper we demonstrate that many of the mathematical relationships for
entropies generalize to the quantum spectral information rates.

\section{Preliminaries}

\subsection{Spectral Projections}

The quantum information spectrum approach requires the extensive use of
spectral operators.  For a self-adjoint operator $A$ written in its spectral
decomposition $A = \sum_i \lambda_i |i\rangle \langle i|$ we define the
positive spectral projection on $A$ as
\begin{equation}
\{ A \geq 0 \} = \sum_{\lambda_i \geq 0} |i\rangle \langle i|
\end{equation}
the projector onto the eigenspace of positive eigenvalues of $A$.
Corresponding definitions apply for the other spectral projections $\{ A < 0
\}$, $\{ A > 0 \}$ and $\{ A \leq 0 \}$.  For two operators $A$ and $B$, we can
then define $\{ A \geq B \}$ as $\{ A - B \geq 0 \}$, and similarly for the
other ordering relations.

\subsection{Two Important Lemmas}

Here the two key lemmas for many results in this paper are presented.
\begin{lemma}
\label{lemma}
For self-adjoint operators $A$, $B$ and any positive operator $0 \leq P \leq I$
the inequality
\begin{equation}
\mathrm{Tr}\big[ P(A-B)\big] \leq \mathrm{Tr}\big[ \big\{ A \geq B \big\}
(A-B)\big]
\label{eqn:first_ineq}
\end{equation}
holds.
\end{lemma}
\begin{proof}
As both operators $A$ and $B$ are self-adjoint so is their difference $A - B$.
Hence, we can diagonalize $A - B$ and write it as the difference of two
positive diagonal operators $\Pi \geq 0$ and $\Omega \geq 0$, such that $U(A -
B)U^{\dag} = \Pi - \Omega$.  As the operator $\big\{ A \geq B \big\}$ projects
onto the positive eigenvalues of $A - B$, we find
\begin{equation}
\mathrm{Tr}\big[ \big\{ A \geq B \big\} (A-B)\big] = \mathrm{Tr} \big[ \Pi
\big]
\label{eqn:proj_ineq1}
\end{equation}
For any positive $P \leq I$ we then have
\begin{align}
\mathrm{Tr}\big[ P(A-B) ] &= \mathrm{Tr}\big[ P'(\Pi - \Omega) \big] \nonumber
\\
&= \mathrm{Tr}\big[P'\Pi \big] - \mathrm{Tr}\big[P'\Omega \big] \nonumber \\
&\leq \mathrm{Tr}\big[\Pi \big]
\label{eqn:proj_ineq2}
\end{align}
where $P' = UPU^{\dag} \leq I$ is positive.  Combining (\ref{eqn:proj_ineq1})
with (\ref{eqn:proj_ineq2}) gives the required inequality in
(\ref{eqn:first_ineq}).
\end{proof}

\begin{lemma}
\label{lemma2}
For self-adjoint operators $A$ and $B$, and any completely positive
trace-preserving (CPTP) map $\mathcal{T}$ the inequality
\begin{equation}
\mathrm{Tr}\big[ \{\mathcal{T}(A) \geq \mathcal{T}(B) \}\mathcal{T}(A-B)\big]
\leq \mathrm{Tr}\big[ \big\{ A \geq B \big\} (A-B)\big]
\label{eqn:second_ineq}
\end{equation}
holds.
\end{lemma}
\begin{proof}
The operator $A-B$ may be expressed in terms of a difference of two diagonal
positive operators $U(A-B)U^{\dag} = \Pi - \Omega$.  As $\mathcal{T}$ is a CPTP
map it may be written in a Kraus representation as $\mathcal{T}(A) = \sum_k T_k
A T_k^{\dag}$.  Defining $\mathcal{T}'$ by the Kraus operators $T'_k = T_k
U^{\dag}$ implies that $\mathcal{T}'$ is also a CPTP map.  Define $\Delta =
\mathrm{Tr}\big[ \{\mathcal{T}(A) \geq \mathcal{T}(B) \}\mathcal{T}(A-B)\big] -
\mathrm{Tr}\big[ \big\{ A \geq B \big\} (A-B)\big]$, then
\begin{align}
\Delta &= \mathrm{Tr}\big[ \{\mathcal{T}(A) \geq \mathcal{T}(B)
\}\mathcal{T}(A-B)\big] - \mathrm{Tr}\big[ \Pi \big] \\
&\leq \mathrm{Tr}\big[ \{\mathcal{T}(A) \geq \mathcal{T}(B)
\}\mathcal{T}'(\Pi)\big] - \mathrm{Tr}\big[ \Pi \big]\label{eqn:positive} \\
&\leq \mathrm{Tr}\big[ \mathcal{T}'(\Pi) \big] - \mathrm{Tr}\big[ \Pi \big] \\
&= 0 \label{eqn:trace_preserving}
\end{align}
where (\ref{eqn:positive}) follows from $\mathcal{T}'$ being a completely
positive map, implying that $\mathcal{T}'(\Pi)$ and $\mathcal{T}'(\Omega)$ are
both positive, and (\ref{eqn:trace_preserving}) is due to $\mathcal{T}'$ being
a trace-preserving map.
\end{proof}

\section{Quantum Spectral Divergence Rates}

The spectral divergence rates act as generalizations of the relative entropy.
They are defined on sequences of states $\rho = \{ \rho_n \}_{n=1}^{\infty}$
(and operators), unlike the relative entropy which is defined for individual
states (and operators).
\begin{definition}
For sequences of states $\rho = \{ \rho_n \}_{n=1}^{\infty}$ and positive
operators $\omega = \{ \omega_n \}_{n=1}^{\infty}$, define the difference operator $\Pi_n(\gamma) =
\rho_n - e^{n\gamma}\omega_n$, then the quantum spectral sup-(inf-)divergence
rates are defined as
\begin{align}
\overline{D}(\rho \| \omega) &= \inf \Big\{ \gamma : \lim_{n\rightarrow \infty}
\mathrm{Tr}\big[ \{ \Pi_n(\gamma) \geq 0 \} \Pi_n(\gamma) \big] = 0 \Big\}
\label{eqn:sup_div} \\
\underline{D}(\rho \| \omega) &= \sup \Big\{ \gamma : \lim_{n\rightarrow
\infty} \mathrm{Tr}\big[ \{ \Pi_n(\gamma) \geq 0 \} \Pi_n(\gamma) \big] = 1
\Big\} \label{eqn:inf_div}
\end{align}
respectively.
\end{definition}

Although the use of sequences allows for immense freedom in choosing them,
there remain a number of basic properties of the quantum spectral divergence
rates that hold for all sequences.  In the {i.i.d.} case the sequence is
generated from product states $\rho = \{ \varrho^{\otimes n}
\}_{n=1}^{\infty}$, which is used to relate the spectral entropy rates for the
sequence $\rho$ to the entropy of a single state $\varrho$.

\subsection{Equivalence to Previous Definitions}

Although the definitions for the spectral divergences differ slightly from
those in (38) and (39) of \cite{hayashi03}, they are equivalent, as the next
propositions show.
\begin{proposition}
\label{equiv_sup}
The spectral sup-divergence rate $\overline{D}(\rho\| \omega)$ is equal to
\begin{equation}
\overline{\mathcal{D}}(\rho\| \omega) = \inf \Big\{ \alpha : \lim_{n\rightarrow
\infty} \mathrm{Tr}\big[ \{ \rho_n \geq e^{n\alpha}\omega_n \} \rho_n \big] = 0
\Big\}
\end{equation}
which is the previously used definition of the spectral sup-divergence rate.
Hence the two definitions are equivalent.
\end{proposition}
\begin{proof}
For any $\alpha = \overline{\mathcal{D}}(\rho\| \omega) + \delta$, with $\delta
> 0$, implies
\begin{align}
0 &= \lim_{n\rightarrow \infty} \mathrm{Tr}\big[ \{ \rho_n \geq
e^{n\alpha}\omega_n \} \rho_n \big] \nonumber \\
&\geq \lim_{n\rightarrow \infty} \mathrm{Tr}\big[ \{ \rho_n \geq
e^{n\alpha}\omega_n \} (\rho_n - e^{n\alpha}\omega_n) \big] \nonumber \\
&\geq 0
\end{align}
giving $\overline{\mathcal{D}}(\rho\| \omega) \geq \overline{D}(\rho\|
\omega)$, as $\delta$ is arbitrary.  For the converse we assume that the
inequality is strict, such that $\overline{\mathcal{D}}(\rho\| \omega) =
\overline{D}(\rho\| \omega) + 4\delta$ for some $\delta > 0$.  Then choosing
$\alpha = \overline{D}(\rho\| \omega) + 2\delta$, $\gamma = \overline{D}(\rho\|
\omega) + \delta$, we have from Lemma \ref{lemma},
\begin{align}
\mathrm{Tr}\big[ \{ \rho_n \geq e^{n\alpha}\omega_n \} \rho_n \big] &\leq
\mathrm{Tr}\big[ \{ \rho_n \geq e^{n\gamma}\omega_n \} (\rho_n -
e^{n\gamma}\omega_n) \big] \nonumber \\
&\phantom{=}\: + e^{n\gamma}\mathrm{Tr}\big[ \{ \rho_n \geq e^{n\alpha}\omega_n
\} \omega_n \big] \nonumber \\
&\leq \epsilon_n + e^{-n\delta}
\end{align}
where $\epsilon_n = \mathrm{Tr}\big[ \{ \rho_n \geq e^{n\gamma}\omega_n \}
(\rho_n - e^{n\gamma}\omega_n) \big]$ and $\mathrm{Tr}\big[ \{ \rho_n \geq
e^{n\alpha}\omega_n \}\omega_n \big] \leq e^{-n\alpha}$ holds for any $\alpha$.
As the right hand side goes to zero asymptotically and since $\alpha <
\overline{\mathcal{D}}(\rho\| \omega)$ we have a contradiction.
\end{proof}
\begin{proposition}
\label{equiv_inf}
The spectral inf-divergence rate $\underline{D}(\rho\| \omega)$ is equivalent
to
\begin{equation}
\underline{\mathcal{D}}(\rho\| \omega) = \sup \Big\{ \alpha :
\lim_{n\rightarrow \infty} \mathrm{Tr}\big[ \{ \rho_n \geq e^{n\alpha}\omega_n
\} \rho_n \big] = 1 \Big\}
\end{equation}
which is the previously used definition of the spectral inf-divergence rate.
\end{proposition}
\begin{proof}
For any $\alpha = \underline{D}(\rho\| \omega) - \delta$, with $\delta > 0$,
implies
\begin{align}
1 &\geq \lim_{n\rightarrow \infty} \mathrm{Tr}\big[ \{ \rho_n \geq
e^{n\alpha}\omega_n \} \rho_n \big] \nonumber \\
&\geq \lim_{n\rightarrow \infty} \mathrm{Tr}\big[ \{ \rho_n \geq
e^{n\alpha}\omega_n \} (\rho_n - e^{n\alpha}\omega_n) \big] \nonumber \\
&= 1
\end{align}
giving $\underline{\mathcal{D}}(\rho\| \omega) \geq \underline{D}(\rho\|
\omega)$, as $\delta$ is arbitrary.  For the converse we assume that the
inequality is strict, such that $\underline{\mathcal{D}}(\rho\| \omega) =
\underline{D}(\rho\| \omega) + 4\delta$ for some $\delta > 0$.  Then choosing
$\alpha = \underline{\mathcal{D}}(\rho\| \omega) - \delta$, $\gamma =
\underline{\mathcal{D}}(\rho\| \omega) - 2\delta$, we have from Lemma
\ref{lemma},
\begin{align}
1 &\overset{n\rightarrow \infty}\leftarrow \mathrm{Tr}\big[ \{ \rho_n \geq
e^{n\alpha}\omega_n \} \rho_n \big] \nonumber \\
&\leq \mathrm{Tr}\big[ \{ \rho_n \geq e^{n\gamma}\omega_n \} (\rho_n -
e^{n\gamma}\omega_n) \big] \nonumber \\
&\phantom{=}\: + e^{n\gamma}\mathrm{Tr}\big[ \{ \rho_n \geq e^{n\alpha}\omega_n
\} \omega_n \big] \nonumber \\
&\leq \mathrm{Tr}\big[ \{ \rho_n \geq e^{n\gamma}\omega_n \} (\rho_n -
e^{n\gamma}\omega_n) \big] + e^{-n\delta}
\end{align}
where $\mathrm{Tr}\big[ \{ \rho_n \geq e^{n\alpha}\omega_n \}\omega_n \big]
\leq e^{-n\alpha}$ holds for any $\alpha$. Thus $\lim_{n\rightarrow \infty}
\mathrm{Tr}\big[ \{ \rho_n \geq e^{n\gamma}\omega_n \} (\rho_n -
e^{n\gamma}\omega_n) \big] = 1$, where $\gamma > \underline{D}(\rho\|\omega)$,
which is a contradiction.
\end{proof}

Despite the above equivalences, it is useful to use the definitions in
(\ref{eqn:sup_div}) and (\ref{eqn:inf_div}) for the divergence rates as they
allow the application of Lemmas \ref{lemma} and \ref{lemma2} in deriving
various properties of these rates.

\subsection{Properties of Spectral Divergences}

\begin{proposition}
The spectral divergence rates for a sequence of states $\rho = \{ \rho_n \}_{n
= 1}^{\infty}$ are related by
\begin{equation}
\underline{D}(\rho \| \omega ) \leq \overline{D}(\rho \| \omega )
\end{equation}
for any positive sequence of operators $\omega = \{ \omega_n \}_{n =
1}^{\infty}$.
\end{proposition}
\begin{proof}
Let $\gamma$ be any real number such that
\begin{equation}
\lim_{n\rightarrow \infty} \mathrm{Tr}\big[ \{ \rho_n \geq e^{n\gamma}\omega_n
\} (\rho_n - e^{n\gamma}\omega_n) \big] = 0
\end{equation}
then for any $\alpha = \gamma + \delta$, for $\delta > 0$, we have from Lemma
\ref{lemma}
\begin{align}
\mathrm{Tr}\big[ \{ \rho_n \geq e^{n\alpha}\omega_n \} \rho_n \big] &\leq
\mathrm{Tr}\big[ \{ \rho_n \geq e^{n\gamma}\omega_n \} (\rho_n -
e^{n\gamma}\omega_n) \big] \nonumber \\
&\phantom{=}\: + e^{n\gamma}\mathrm{Tr}\big[ \{ \rho_n \geq e^{n\alpha}\omega_n
\} \omega_n \big] \nonumber \\
&\leq \mathrm{Tr}\big[ \{ \rho_n \geq e^{n\gamma}\omega_n \} (\rho_n -
e^{n\gamma}\omega_n) \big] \nonumber \\
&\phantom{=}\: + e^{-n\delta}
\end{align}
and the right hand side goes to zero asymptotically.  Hence
\begin{equation}
\lim_{n\rightarrow \infty} \mathrm{Tr}\big[ \{ \rho_n \geq e^{n\alpha}\omega_n
\} (\rho_n - e^{n\alpha}\omega_n) \big] = 0
\end{equation}
for any $\alpha \geq \gamma$.
\end{proof}

\begin{proposition}
\label{div_decrease_cptp}
Under any sequence of CPTP maps $\mathcal{T} = \{ T_n \}_{n=1}^{\infty}$ the
spectral divergence rates can only decrease, that is
\begin{align}
\overline{D}(\rho \| \omega) \geq \overline{D}(\mathcal{T}(\rho) \|
\mathcal{T}(\omega)) \\
\underline{D}(\rho \| \omega) \geq \underline{D}(\mathcal{T}(\rho) \|
\mathcal{T}(\omega))
\end{align}
in analogy with the monotonicity of the quantum relative entropy.
\end{proposition}
\begin{proof}
For any $\delta > 0$ choose $\gamma = \overline{D}(\rho\| \omega) + \delta$,
then from Lemma \ref{lemma2} we have
\begin{align}
0 &\leq \mathrm{Tr}\big[ \{\mathcal{T}(\rho_n) \geq
e^{n\gamma}\mathcal{T}(\omega_n) \}\mathcal{T}(\rho_n -
e^{n\gamma}\omega_n)\big] \\
&\leq \mathrm{Tr}\big[ \big\{ \rho_n \geq e^{n\gamma}\omega_n \big\} (\rho_n -
e^{n\gamma}\omega_n)\big] \\
&\overset{n\rightarrow \infty}\rightarrow 0
\end{align}
and hence $\overline{D}(\mathcal{T}(\rho)\|\mathcal{T}(\omega)) \leq
\overline{D}(\rho\|\omega) + \delta$ for all $\delta > 0$ implying the
inequality holds.

Similarly, choose $\gamma =
\underline{D}(\mathcal{T}(\rho)\|\mathcal{T}(\omega)) - \delta$, then from
Lemma \ref{lemma2} we have
\begin{align}
1 &\overset{n\rightarrow \infty}\leftarrow \mathrm{Tr}\big[
\{\mathcal{T}(\rho_n) \geq e^{n\gamma}\mathcal{T}(\omega_n)
\}\mathcal{T}(\rho_n - e^{n\gamma}\omega_n)\big] \\
&\leq \mathrm{Tr}\big[ \big\{ \rho_n \geq e^{n\gamma}\omega_n \big\} (\rho_n -
e^{n\gamma}\omega_n)\big] \\
&\leq 1
\end{align}
and hence $\underline{D}(\mathcal{T}(\rho)\|\mathcal{T}(\omega)) \leq
\underline{D}(\rho \| \omega) + \delta$ for all $\delta > 0$.
\end{proof}

\begin{corollary}
The spectral divergence rates between two sequences of states are non-negative.
\end{corollary}
\begin{proof}
Choose $\mathcal{T}$ to be the trace operation.  Then for any $\gamma < 0$ we
have $\lim_{n\rightarrow \infty} \{ 1 \geq e^{n\gamma}\}(1 - e^{n\gamma}) = 1$
and hence $\overline{D}(\rho \| \omega) \geq \underline{D}(\rho \| \omega) \geq
0$.
\end{proof}

Note that the spectral divergence rates between operators can be negative.  An
example of this that is introduced later is the conditional spectral entropy
rates, which can be either positive or negative, and these are defined in terms
of the divergence rates between the sequence of bipartite states and a sequence
of operators derived from those states.

\section{Spectral Information Rates}

Spectral information rates, the generalizations of entropy, conditional entropy
and mutual information, may be defined in terms of the spectral divergence
rates.  In this section, the properties of the spectral information rates are
examined and their relationship to the properties of the corresponding entropic
quantities discussed.

\subsection{Spectral Entropy Rates}
\begin{definition}
The sup-spectral entropy rate is defined for a sequence of states $\rho = \{
\rho^{X}_n\}_{n=1}^{\infty}$ of a quantum system $X$ as
\begin{equation}
\overline{S}(X) = -\underline{D}(\rho \| I)
\end{equation}
where $I = \{ I^X_n \}_{n=1}^{\infty}$.  The inf-spectral entropy rate
$\underline{S}(X)$, is defined as
\begin{equation}
\underline{S}(X) = -\overline{D}(\rho \| I)
\end{equation}
for a given sequence.
\end{definition}

The spectral entropy rates defined here are equivalent to the quantities
obtained from the definitions in \cite{nagaoka02}, which can be shown in a
similar way to Propositions \ref{equiv_sup} and \ref{equiv_inf}.

\begin{proposition}
\label{ent_bounds}
The spectral entropy rates are bounded above and below by
\begin{equation}
0 \leq \underline{S}(X) \leq \overline{S}(X) \leq \log d
\end{equation}
where the Hilbert space $\mathcal{H}_n$ of the system $X$ is of dimension
$d^n$.
\end{proposition}
\begin{proof}
For any $\gamma > 0$ the spectral projection $\{\rho_n \geq e^{n\gamma} \} = 0$
as $e^{n\gamma} > 1 \geq \lambda$ for $\lambda$ any eigenvalue of $\rho_n$.
Hence,
\begin{equation}
\underline{D}(\rho \| I ) \leq \overline{D}(\rho \| I ) \leq 0
\end{equation}
and thus $\overline{S}(X) \geq \underline{S}(X) \geq 0$.

To show $\overline{S}(X) \leq \log d$, we have
\begin{equation}
0 \leq \underline{D}(\rho \| e^{-n \log d}I ) = \log d - \overline{S}(X)
\end{equation}
and hence $\underline{S}(X) \leq \overline{S}(X) \leq \log d$.
\end{proof}

The next proposition states that any sequence of complete measurements on a
system increases the spectral entropy rates.  This property is the direct
analogue of the {i.i.d.} case (in which the spectral entropy reduces to the von
Neumann entropy).  All complete measurements on a system are represented by
unital CPTP maps on the system, assuming no conditioning on the result.
\begin{proposition}
For any sequence of unital CPTP maps $\mathcal{T}$ and sequence of states
$\rho$ the inequalities
\begin{align}
\underline{S}(\mathcal{T}(X)) &\geq \underline{S}(X) \\
\overline{S}(\mathcal{T}(X)) &\geq \overline{S}(X)
\end{align}
both hold.
\end{proposition}
\begin{proof}
From the definitions of the spectral entropy rates, and using Proposition
\ref{div_decrease_cptp}
\begin{align}
\underline{S}(\mathcal{T}(X)) &= -\overline{D}(\mathcal{T}(\rho)\|I) =
-\overline{D}(\mathcal{T}(\rho)\|\mathcal{T}(I)) \nonumber \\
&\geq -\overline{D}(\rho\|I) = \underline{S}(X)
\end{align}
where $\mathcal{T}(I) = I$ as $T_n$ is unital for all $n$.  The proof for the
sup-spectral entropy rate is similar.
\end{proof}

It may be noted that for bipartite sequences of pure states $\rho^{AB} = \{
|\phi^{AB}\rangle \langle \phi^{AB}|_n \}_{n=1}^{\infty}$ the reduced states
$\rho^A_n$ and $\rho^B_n$ have identical spectra.  Hence it is immediate that
the spectral entropy rates for the reduced states are equal
\begin{align}
\overline{S}(A) &= \overline{S}(B) \label{eqn:pure_eq1} \\
\underline{S}(A) &= \underline{S}(B) \label{eqn:pure_eq2}
\end{align}
for sequences of bipartite pure states.

\subsection{Spectral Conditional Entropy Rates}

\begin{definition}
The spectral conditional entropy rates for sequences of bipartite states are
defined as
\begin{equation}
\overline{S}(A|B) = -\underline{D}(\rho^{AB} \| I^A \otimes \rho^B)
\end{equation}
and
\begin{equation}
\underline{S}(A|B) = -\overline{D}(\rho^{AB} \| I^A \otimes \rho^B)
\end{equation}
respectively.
\end{definition}

Next, we give a relationship showing that the conditional spectral entropy
rates are necessarily less than the corresponding spectral entropy rate of a
source.
\begin{proposition}
\label{cond_dec}
Conditioning reduces the spectral entropy rate, such that
\begin{align}
\overline{S}(A|BC) &\leq \overline{S}(A|B) \leq \overline{S}(A) \\
\underline{S}(A|BC) &\leq \underline{S}(A|B) \leq \underline{S}(A)
\end{align}
for any tripartite sequence $\rho^{ABC} = \{\rho^{ABC}_n\}_{n=1}^{\infty}$.
\end{proposition}
\begin{proof}
The inequalities follow from Proposition \ref{div_decrease_cptp} and the fact
that the partial trace is a CPTP map.
\end{proof}

The \textit{chain rules} \cite{cover} in information theory relate the
entropies, $H(X)$ and $H(XY)$, to the conditional entropy $H(Y|X)$ and mutual
information $I(X:Y)$, {e.g.} $H(XY) = H(X) + H(Y|X)$.  Although the equalities
given for the various chain rules do not hold in general, the spectral
information rates are related by sets of inequalities.  Examples are known, in
each case, where the inequality is strict.
\begin{proposition}
\label{chain1}
For sequences of bipartite states the conditional spectral entropy is related to the
spectral entropies by
\begin{equation}
\underline{S}(A|B) \geq \underline{S}(AB) - \overline{S}(B)
\end{equation}
giving a chain rule inequality.
\end{proposition}
%%%%%%%%%%%%%% NEW PROOF %%%%%%%%%%%%%
\begin{proof}
Defining the difference operators $\Pi_n(\alpha - \beta) = \rho_n^{AB} - e^{-n(\alpha - \beta)} I_n^A\otimes \rho_n^B$, and the projections $P_1 =  \{ \rho_n^{AB} \geq e^{-n(\alpha - \beta)} I_n^A\otimes \rho_n^B \}$, $P_2 = I_n^A \otimes \{ \rho_n^B \geq e^{-n\beta} \}$ and $\overline{P}_2 = 1 - P_2$, we have
\begin{align}
0 &\leq \mathrm{Tr}\big[ P_1 \Pi_n(\alpha - \beta) \big] \nonumber \\
&= \mathrm{Tr}\big[P_1(P_2+\overline{P}_2) \Pi_n(\alpha - \beta) (P_2+\overline{P}_2)\big] \nonumber \\
&= \mathrm{Tr}\big[ P_1 P_2 \Pi_n(\alpha - \beta) P_2 \big] + \mathrm{Tr}\big[P_1 \overline{P}_2 \Pi_n(\alpha - \beta) \overline{P}_2 \big] \nonumber \\
&\phantom{=}\; + \mathrm{Tr}\big[P_1 P_2\rho_n^{AB} \overline{P}_2  + \overline{P}_2 \rho_n^{AB} P_2 P_1 \big] \label{eqn:chain1} \\
&\leq \mathrm{Tr}\big[\{ \rho_n^{AB} \geq e^{-n\alpha} \} \big( \rho_n^{AB} - e^{-n\alpha} I_n^{AB}\big) ] \nonumber \\
&\phantom{=}\;+ \mathrm{Tr}[\{ \rho_n^{B} < e^{-n\beta} \} \rho_n^{B} ] \nonumber \\
&\phantom{=}\;+ 2\sqrt{\mathrm{Tr}\big[ \{ \rho_n^{B} < e^{-n\beta} \}\rho_n^{B}\big]\cdot \mathrm{Tr}\big[P_1 P_2 \rho_n^{AB} P_2\big]} \label{eqn:chain2}
\end{align}
The expression in (\ref{eqn:chain1}) is obtained by noting that as $P_2$ and $\overline{P}_2$ both commute with $I^A\otimes \rho^B_n$, the cross-terms in $P_2$ and $\overline{P}_2$ vanish.  The final term in (\ref{eqn:chain2}) is obtained as follows. Using the cyclicity of
the trace, we can write
\begin{equation}
\mathrm{Tr}\big[P_1 P_2\rho_n^{AB} \overline{P}_2  + \overline{P}_2 \rho_n^{AB} P_2 P_1 \big] = \mathrm{Tr}\big[ B^\dagger A + A^\dagger B \big]
\end{equation}
where $A:={\sqrt{\rho_n^{AB}}}\,\overline{P}_2$ and $B:=
{\sqrt{\rho_n^{AB}}}P_2 P_1$. Since the operator $(A^\dagger B + B^\dagger
A\big)$ is self--adjoint,
\begin{align}
\big(\mathrm{Tr}\big[ B^\dagger A + A^\dagger B \big]\big)^2
&= 4 \big(\mathrm{Re}\,\mathrm{Tr}\big[A^\dagger B \big]\big)^2\nonumber\\
&\leq 4 |\mathrm{Tr}\big(A^\dagger B)|^2 \nonumber \\
&\leq 4 \mathrm{Tr}[A^{\dag}A]\cdot
\mathrm{Tr}[B^{\dag}B].
\label{460}
\end{align}
where the last inequality is the Cauchy-Schwarz inequality for the Hilbert-Schmidt
inner product.  We then utilize the fact that
$\mathrm{Tr}[A^{\dag}A] = \mathrm{Tr}[ \overline{P}_2\,
\rho_n^{AB}\overline{P}_2]
=  \mathrm{Tr}[\{\rho^B_n < e^{-n\beta}\}\rho^B_n]$ to obtain the resultant inequality.

Choosing $\alpha = \underline{S}(AB) - \delta$ and $\beta = \overline{S}(B) + \delta$ for arbitrary $\delta > 0$ implies that all the terms in inequality (\ref{eqn:chain2}) vanish in the limit as $n\rightarrow \infty$.  Hence, we have
\begin{equation}
\underline{S}(A|B) \geq \underline{S}(AB) - \overline{S}(B) - 2\delta
\end{equation}
for all $\delta > 0$.
\end{proof}
%%%%%%%%%%%%%%% END NEW PROOF %%%%%%%%%%%%%%%%

\begin{corollary}
\label{chain2}
For sequences of bipartite states the conditional spectral entropy is related to the
spectral entropies by
%%%% begin insertion %%%%%%%%%
\begin{equation}
\overline{S}(A|B) \geq \max \big[\overline{S}(AB) - \overline{S}(B),
 \underline{S}(AB) - \underline{S}(B)\big] \label{eqn:chain_over}
\end{equation}
giving further chain rule inequalities.
\end{corollary}
\begin{proof}
To obtain the first inequality in (\ref{eqn:chain_over}), simply substitute $\alpha = \overline{S}(A|B) + \overline{S}(B) + 2\delta$ and $\beta
= \overline{S}(B) + \delta$ for arbitrary $\delta > 0$ into the proof
of Proposition \ref{chain1}.

For the second inequality in (\ref{eqn:chain_over}) we bound the term
\begin{align}
\mathrm{Tr}\big[P_1 P_2 \rho_n^{AB} P_2\big] &= \mathrm{Tr}\big[P_1 P_2 (\rho_n^{AB} - e^{-n\gamma}I_n^A \otimes \rho_n^B) P_2\big] \nonumber \\
&\phantom{=}\;+ e^{-n\gamma}\mathrm{Tr}\big[P_1 P_2(I_n^A \otimes \rho_n^B)P_2\big].
\nonumber\\
&\leq \mathrm{Tr}\big[\{ \rho_n^{AB} \geq e^{-n\alpha} \} \big(
\rho_n^{AB} - e^{-n\alpha} I_n^{AB}\big) ] \nonumber \\
&\phantom{=}\;+ e^{-n\gamma}\mathrm{Tr}\big[P_1 (I_n^A \otimes \rho_n^B)P_1P_2\big]
\nonumber\\
&\phantom{=}\;+ e^{-n\gamma}\mathrm{Tr}\big[P_1 (I_n^A \otimes
\rho_n^B){\overline{P}}_1P_2\big]
\end{align}
using Lemma \ref{lemma}, where $\gamma = \alpha - \beta$.  The last two terms are obtained by noting that $P_2$ commutes with $I_n^A \otimes \rho_n^B$, and $P_1 + \overline{P}_1 = I_n^{AB}$.  Substituting this relation into the final term of (\ref{eqn:chain2}), and choosing $\beta = \underline{S}(AB) - \overline{S}(A|B) - 2\delta$ and $\alpha = \underline{S}(AB) - \delta$, for an arbitrary $\delta >0$, we have $\mathrm{Tr}\big[P_1
\Pi_n(\alpha - \beta)\big] \rightarrow 1$ as $n \rightarrow \infty$, and hence
$\mathrm{Tr}\big[P_1 (e^{-n\gamma}I_n^A \otimes \rho_n^B) P_1P_2\big] \rightarrow 0$
in this limit. Moreover, since $\mathrm{Tr}\big[P_1\Pi_n(\gamma){\overline{P}}_1\big]=0$,
we have,
\begin{align}
\big|e^{-n\gamma}\mathrm{Tr}\big[P_1 (I_n^A \otimes \rho_n^B){\overline{P}}_1P_2
\big]\big| &= \big|\mathrm{Tr}[P_1 \rho_n^{AB}{\overline{P}}_1P_2
]\big| \nonumber\\
&\leq \sqrt{{\mathrm{Tr}}[P_2P_1\rho P_1] \mathrm{Tr}[{\overline{P}}_1 \rho_n^{AB}]}\nonumber\\
&\rightarrow 0 \,\, {\hbox{as }} n \rightarrow \infty,
\end{align}
whenever $\gamma > {\overline{S}}(A|B)$, as $\mathrm{Tr}\big[
{\overline{P}}_1 \rho_n^{AB}\big] \rightarrow 0$.  Hence, the first and third terms of the sum in (\ref{eqn:chain2}) vanish asymptotically, and therefore $\underline{S}(B) \geq \underline{S}(AB) - \overline{S}(A|B) - 2\delta$ for all $\delta > 0$.
\end{proof}

\begin{proposition}
\label{chain3}
For bipartite states the conditional spectral entropy is related to the
spectral entropies by
\begin{equation}
\overline{S}(A|B) \leq \overline{S}(AB) - \underline{S}(B)
\end{equation}
giving a chain rule inequality.
\end{proposition}
\begin{proof}
%% begin insertion  %%%%
Defining the difference operators $\Pi_n(\alpha + \beta) = \rho_n^{AB} -
e^{-n(\alpha+\beta)} I_n^{AB}$,
the projections $P_1 =  \{ \rho_n^{AB} \geq e^{-n{(\alpha + \beta)}} I_n^{AB}
\}$,
$P_2 = I_n^A \otimes \{ \rho_n^B \geq e^{-n\beta} \}$ and $\overline{P}_2 = 1 -
P_2$, we have
\begin{align}
0 &\leq \mathrm{Tr}\big[ P_1 \Pi_n(\alpha + \beta) \big] \nonumber \\
&= \mathrm{Tr}\big[P_1(P_2+\overline{P}_2) \Pi_n(\alpha + \beta) (P_2+\overline{P}_2)\big] \nonumber \\
&= \mathrm{Tr}\big[ P_1 P_2 \Pi_n(\alpha + \beta) P_2 \big] + \mathrm{Tr}\big[P_1 \overline{P}_2 \Pi_n(\alpha + \beta) \overline{P}_2 \big] \nonumber \\
&\phantom{=}\; + \mathrm{Tr}\big[P_1 \overline{P}_2 \rho_n^{AB} P_2 + P_2 \rho_n^{AB} \overline{P}_2 P_1 \big] \nonumber \\
&\leq \mathrm{Tr}[\{ \rho_n^{B} \geq e^{-n\beta} \} \rho_n^{B} ] \nonumber \\
&\phantom{=}\;+ \mathrm{Tr}\big[\{ \rho_n^{AB} \geq e^{-n\alpha}I_n^{A}\otimes \rho_n^B \} \big( \rho_n^{AB} - e^{-n\alpha} I_n^{A}\otimes \rho_n^B \big) ] \nonumber \\
&\phantom{=}\;+ 2\sqrt{\mathrm{Tr}\big[ \{ \rho_n^{B} \geq e^{-n\beta} \}\rho_n^{B}\big]\cdot \mathrm{Tr}\big[P_1 \overline{P}_2 \rho_n^{AB} \overline{P}_2\big]} \label{eqn:chain3}
\end{align}
proceeding analogously to the proof of Proposition \ref{chain1}. Choosing
$\alpha = \overline{S}(AB) - \underline{S}(B) + 2\delta$ and $\beta =
\underline{S}(B) - \delta$ for arbitrary $\delta > 0$ implies the required
inequality in the limit $n\rightarrow \infty$.
\end{proof}
%% end insertion %%
\begin{corollary}
\label{chain4}
For sequences of bipartite states the conditional spectral entropy is related to the
spectral entropies by
\begin{equation}
\underline{S}(A|B) \leq \min \big[\underline{S}(AB) - \underline{S}(B),
 \overline{S}(AB) - \overline{S}(B)\big]
\end{equation}
giving further chain rule inequalities.
\end{corollary}
\begin{proof}
The first inequality is obtained by substituting $\alpha = \underline{S}(A|B) - \delta$ and $\beta = \underline{S}(B)
- \delta$ into the proof of Proposition \ref{chain3}.

For the second inequality note that
\begin{align}
&\mathrm{Tr}\big[P_1 {\overline{P}}_2 \rho_n^{AB} {\overline{P}}_2 \big]
\nonumber\\
&= \mathrm{Tr}\big[P_1 {\overline{P}}_2 (\rho_n^{AB} - e^{-n(\alpha + \beta)
}I_n^A \otimes I_n^B){\overline{P}}_2 \big]
\nonumber\\
&\phantom{=}\;+ e^{-n(\alpha + \beta)}\mathrm{Tr}\big[P_1 {\overline{P}}_2(I_n^A
\otimes I_n^B){\overline{P}}_2\big]
\nonumber\\
&\leq \mathrm{Tr}\big[\{ {\overline{P}}_2 P_1 {\overline{P}}_2 \big(
\rho_n^{AB} - e^{-n\alpha} I_n^{A} \otimes \rho_n^B \big)\big] \nonumber \\
&\phantom{=}\;+ e^{-n(\alpha + \beta)}\mathrm{Tr}\big[\overline{P}_2 P_1 \overline{P}_2\big]
\nonumber\\
&\leq \mathrm{Tr}\big[\{ \rho_n^{AB} \geq e^{-n\alpha}I_n^{A} \otimes \rho_n^B
\} \big( \rho_n^{AB} - e^{-n\alpha} I_n^{A} \otimes \rho_n^B \big)
\big]\nonumber \\
&\phantom{=}\;+ \mathrm{Tr}\big[P_1(e^{-n(\alpha + \beta)}I_n^{AB})\big].
\label{chain557}
\end{align}
and
\begin{equation}
\mathrm{Tr}\big[P_2\rho_n^{AB}\big] =
\mathrm{Tr}\big[\{ \rho_n^{B} \geq e^{-n\beta} \}\rho_n^{B}\big]
\label{chain558}
\end{equation}
Substituting (\ref{chain557}) and (\ref{chain558}) into the right hand side of (\ref{eqn:chain3}) and
choosing $\beta = \overline{S}(AB) - {\underline{S}}(A|B) + 2\delta$ and $\alpha =
{\underline{S}}(A|B) - \delta$, for arbitrary $\delta >0$, yields the
desired inequality in the limit $n \rightarrow \infty$. This relies on the fact that for the given values of $\alpha$ and $\beta$ the term
$\mathrm{Tr}\big[P_1\big(e^{-n(\alpha + \beta)}I_n^{AB})\big]$ vanishes in this limit.
\end{proof}

%%%%%%% end insertion %%%%%%%%%%%%%
The chain rule inequalities may then be applied to derive many properties that
are the generalizations of entropic inequalities.

\begin{corollary}
The conditional spectral entropy rates are bounded above and below by
\begin{equation}
-\log d \leq -\overline{S}(A) \leq \underline{S}(A|B) \leq \overline{S}(A|B)
\leq \log d
\end{equation}
for Hilbert spaces $\mathcal{H}^A_n$ of dimension $d^n$.
\end{corollary}
\begin{proof}
For each state $\rho^{AB}_n$ take a purification $|\psi^{ABC}\rangle \langle
\psi^{ABC}|_n$.  From the chain rule inequalities and Proposition
\ref{ent_bounds} it then follows that
\begin{equation}
\underline{S}(A|BC) \leq \underline{S}(A|B) \leq \overline{S}(A|B) \leq
\overline{S}(A) \leq \log d \nonumber \; .
\end{equation}
Using the chain rule inequality, then for states that are purifications on
$ABC$ we have,
\begin{equation}
-\overline{S}(A) = \underline{S}(ABC) - \overline{S}(BC) \leq
\underline{S}(A|BC)
\end{equation}
as $\underline{S}(ABC) = 0$ and $\overline{S}(BC) = \overline{S}(A)$.
\end{proof}

The strong-subadditivity relationships follow immediately from the chain rule
inequalities and the monotonicity of the conditional spectral rates under
partial traces.
\begin{proposition}
The following strong-subadditivity relationships
\begin{align}
\underline{S}(ABC) + \overline{S}(B) &\leq \overline{S}(AB) + \overline{S}(BC)
\\
\overline{S}(ABC) + \underline{S}(B) &\leq \overline{S}(AB) + \overline{S}(BC)
\end{align}
and
\begin{align}
\underline{S}(ABC) + \underline{S}(B) &\leq \overline{S}(AB) +
\underline{S}(BC) \\
\underline{S}(ABC) + \underline{S}(B) &\leq \underline{S}(AB) +
\overline{S}(BC)
\end{align}
hold for all sequences of tripartite states $\rho^{ABC}$.
\end{proposition}
\begin{proof}
These follow from Propositions \ref{chain1} and \ref{chain3}, and their
corollaries, and Proposition \ref{cond_dec}.
\end{proof}

\begin{corollary}
\label{subadd}
The subadditivity relationships
\begin{align}
\overline{S}(AB) &\leq \overline{S}(A) + \overline{S}(B) \\
\underline{S}(AB) &\leq \min \big[ \underline{S}(A) + \overline{S}(B),
\overline{S}(A) + \underline{S}(B) \big]
\end{align}
hold for any sequence of bipartite states.
\end{corollary}
\begin{proof}
For any sequence $\rho^{AB} = \{ \rho^{AB}_n \}_{n=1}^{\infty}$ take a
purification on a system $C$ such that $\rho^{ABC} = \{
|\psi^{ABC}\rangle\langle \psi^{ABC}|_n \}_{n=1}^{\infty}$, and then utilize
strong-subadditivity and the equalities for bipartite pure states given in
(\ref{eqn:pure_eq1}) and (\ref{eqn:pure_eq2}).
\end{proof}

\begin{corollary}
The spectral entropy rates for any bipartite sequence of states obey the
following inequalities,
\begin{align}
\overline{S}(AB) &\geq \big| \overline{S}(A) - \overline{S}(B) \big| \\
\underline{S}(AB) &\geq \max \Big[ \underline{S}(A) - \overline{S}(B),
\underline{S}(B) - \overline{S}(A) \Big]
\end{align}
which are the analogues of the Araki-Lieb inequality \cite{araki70}.
\end{corollary}
\begin{proof}
As for Corollary \ref{subadd}.
\end{proof}

As the quantum information spectrum is a generalization of the classical case,
the properties determined so far also hold for any finite alphabet classical
source.  A classical bipartite source is one where the reduced density matrices
commute with the total state, that is $[ \rho^{AB}, I^A\otimes \rho^B ] = 0$
and $[ \rho^{AB}, \rho^A \otimes I^B] = 0$, where $[\mu, \nu] = \mu \nu - \nu
\mu$ for operators $\mu$ and $\nu$.  For sequences of classical bipartite
states a number of inequalities may be tightened.
\begin{proposition}
The conditional spectral entropy rates are positive for classical states.
\end{proposition}
\begin{proof}
As the states commute we may write them in a common eigenbasis, where
\begin{equation}
\rho^{AB} = \sum_{ij} \lambda_{ij} |ij\rangle\langle ij|_{AB}
\end{equation}
and without loss of generality $I^A\otimes\rho^B = \sum_{ijk} \lambda_{kj}
|ij\rangle\langle ij|_{AB}$.  Therefore we have
\begin{align}
P(\gamma) &= \{ \rho^{AB} \geq e^{-n\gamma}I^A\otimes \rho^B \} \\
&= \Big\{ \sum_{ij} \big(\lambda_{ij} - e^{-n\gamma}\sum_k
\lambda_{kj}\big)|ij\rangle\langle ij| \geq 0 \Big\} \\
&= 0
\end{align}
if $\gamma = -\delta < 0$.  This is due to the fact that $\lambda_{ij} <
e^{n\delta} \sum_k \lambda_{kj}$, for all $i,j$.  Hence we have
$\underline{S}(A|B) \geq 0$.
\end{proof}

\begin{corollary}
For bipartite sequences the following inequalities hold
\begin{align}
\overline{S}(AB) &\geq \max \big[ \overline{S}(A), \overline{S}(B) \big] \\
\underline{S}(AB) &\geq \max \big[ \underline{S}(A), \underline{S}(B) \big]
\end{align}
for all finite-state classical sources.
\end{corollary}

% The chain rule inequalities may be strict, as shown in the following example.
%Take the mixed source
% \begin{equation}
% \rho_n^{AB} = t \frac{1}{d^n} \sum_i |ii\rangle \langle ii|^{AB} + (1-t)
%|00\rangle \langle 00|^{AB}
% \end{equation}
% for $0 < t < 1$, then we have $\underline{S}(AB) - \overline{S}(B) = -\log
%d$, $\underline{S}(A|B) = \underline{S}(AB) - \underline{S}(B) =
%\overline{S}(AB) - \overline{S}(B) = \overline{S}(A|B) = 0$, and
%$\overline{S}(AB)-\underline{S}(B) = \log d$.

\subsection{Spectral Mutual Information Rates}

\begin{definition}
The sup-spectral mutual information rate is defined for a sequence of bipartite
states $\rho^{AB}_n$ as
\begin{equation}
\overline{S}(A:B) = \overline{D}(\rho^{AB}\| \rho^A \otimes \rho^B)
\end{equation}
Similarly, the inf-spectral mutual information rate is defined as
\begin{equation}
\underline{S}(A:B) = \underline{D}(\rho^{AB}\| \rho^A \otimes \rho^B)
\end{equation}
for a given sequence.
\end{definition}

\begin{proposition}
For sequences of bipartite states:
\begin{enumerate}
\item The spectral mutual information rates are always non-negative, $\overline{S}(A:B) \geq \underline{S}(A:B) \geq 0$.
\item The spectral mutual information rates decrease under CPTP mappings on one part of the system.
\item The spectral mutual information rates are monotonic,
\begin{align}
\underline{S}(A:B) &\leq \underline{S}(A:BC) \nonumber \\
\overline{S}(A:B) &\leq \overline{S}(A:BC) \nonumber
\end{align}
under reduction of the system size.
\end{enumerate}
\end{proposition}
\begin{proof}
These properties follow from the definitions and the properties of the spectral divergence rates.
\end{proof}

\begin{proposition}
The following chain rule inequalities hold
\begin{align}
\overline{S}(A:B) &\leq \overline{S}(A) - \underline{S}(A|B) \\
\overline{S}(A:B) &\geq \max \big[ \overline{S}(A) - \overline{S}(A|B), \underline{S}(A) - \underline{S}(A|B) \big] \\
\underline{S}(A:B) &\geq \underline{S}(A) - \overline{S}(A|B) \\
\underline{S}(A:B) &\leq \min \big[ \overline{S}(A) - \overline{S}(A|B), \underline{S}(A) - \underline{S}(A|B) \big]
\end{align}
for sequences of bipartite states.
\end{proposition}
\begin{proof}
The proofs are similar to those given for previous chain rules.
\end{proof}

\section{Discussion}

The general relationships derived here apply to finite state quantum systems,
of which finite alphabet classical states are a subset.  Hence, all the
properties derived apply in standard information theory with the assumption
that the alphabet is finite.  Several results contained in this paper are the finite state quantum
generalizations of the properties described in Theorem 8 of \cite{verdu94}, whilst others represent new inequalities in terms of the information spectrum in classical information theory.

%\bibliographystyle{IEEEtran}
%\bibliography{references}
%\begin{thebibliography}{1}
%H.~Kopka and P.~W. Daly, \emph{A Guide to {\LaTeX}}, 3rd~ed.\hskip 1em plus

%\end{thebibliography}

\end{document}